\documentclass[onecollarge,natbib]{svjour2}
\bibpunct{[}{]}{;}{n}{}{,} 
\smartqed  
\usepackage{slashed}
\usepackage{graphicx}
\usepackage{graphicx}
\usepackage{latexsym}
\newcommand{\ba}{\begin{eqnarray}}
\newcommand{\ea}{\end{eqnarray}}

\usepackage{hyperref}
\newcommand{\nn}{\nonumber}

\newcommand{\bs}{\boldsymbol}
\usepackage{hyperref}

\usepackage{amsmath}
\usepackage{amssymb}
\journalname{Few-Body Systems}
\begin{document}
\title{Wigner distributions of quarks for different polarizations 
\thanks{Presented by Jai More at Light Cone 2016, 5-8 September 2016, IST, Universidade de Lisboa, Portugal.}}


\author{Jai More   \and
        Asmita Mukherjee \and 
          Sreeraj Nair
          }
\institute{Jai More \and Asmita Mukherjee\at
         Indian Institute of Technology, Bombay, Mumbai-400076, India. \\
\email jai.more@phy.iitb.ac.in
           \and
            Asmita Mukherjee  \at \email asmita@phy.iitb.ac.in         \\
                   Sreeraj Nair \at Indian Institute of Science Education and Research, Bhopal, MP-462030, India.\\
                   \email sreeraj\_nair@iiserb.ac.in
}
\date{Received: date / Accepted: date}
\maketitle
\begin{abstract}
We calculate quark Wigner distributions using the light-front
wave functions in a dressed quark model. In this model, a proton target is 
replaced by a simplified spin-1/2 state, namely a quark dressed with a gluon. 
We calculate the Wigner distributions for different polarization configuration 
of quark and the target state in this model. 
\end{abstract}
\section{Introduction}
Generalized parton distributions (GPDs) and transverse momentum dependent parton distributions (TMDs)
\cite{Meissner09} give the three-dimensional picture of the hadronic 
structure. GPDs \cite{Radyushkin97, Ji97} essentially can provide an understanding of the orbital 
angular momentum of partons inside the hadrons. GPDs and TMDs are encompassed in a correlation 
function which describes off-forward  scattering amplitudes that were introduced in Ref. \cite{Meissner08}.
For certain kinematics, if one calculates the matrix element at $z^+=0$ or $z^-=0$, one obtains a reduced 
matrix element called as  generalized transverse momentum dependent parton distributions (GTMDs)
\cite{Meissner09, Meissner08, Lorce13}.  Wigner distributions are obtained by taking Fourier transform of GTMDs 
which are also called as mother distribution functions. The main goal is to understand the partonic 
structure of nucleon that can be achieved by means of position-momentum distributions like Wigner 
distributions. In a particular limit, these distributions can be related to the spin and angular momentum
of quarks and gluons. Wigner distribution is defined as matrix elements between different nucleon states 
as a function of three position and three momentum of a quark or a gluon inside the nucleon in a particular
polarization state. 

Quark and gluon Wigner distribution as a function of three position and 
three momentum variables was first studied by authors in Refs. \cite{Ji03, Belitsky03} in which
relativistic effects were neglected. Wigner distributions \cite{Wigner32} do not give probabilistic interpretation as
they are not positive definite. Only in the classical limit, they become positive definite. 
Inspite of this, a phase-space distribution are worth 
exploring as they can be related to the measurable quantities by integration over transverse position or momentum of the 
parton. The model dependent calculation of Wigner distributions may provide us information about the correlators of quarks
and gluons in a nucleon. Especially, quark orbital angular momentum can be accessed from the Wigner distribution
for unpolarized quarks in the longitudinally polarized nucleon ($F_{14}$) 
\cite{Lorce11, Lorce12, Hagler03, Kanazawa14, Rajan06, Asmita14}. 
These calculations are also a boon in estimating relations of TMDs and GPDs using GTMDs in a certain 
limit. Wigner distributions have been studied in various phenomenological models 
\cite{Lorce12, Asmita14, Asmita15, Pasquini11, Liu14, Liu15, Miller14, Lorce16}.

In this work, we investigate the quark Wigner distributions in light-front Hamiltonian formalism \cite{Hari99}. 
In this formalism, one gets an intuitive picture of deep inelastic scattering for  non-collinear, interacting and 
massive partons. The target state is a dressed quark state and is expanded in terms of light-front wavefunctions 
(LFWFs) which are the multiparton wavefunctions. So, in dressed quark model 
we replace the proton state as a quark dressed with a gluon which incorporates
gluonic degrees of freedom in this perturbative model.

The plan of this paper is as follows. In Sec. \ref{WD}, we calculate the Wigner distribution of quark. 
For different polarization configuration of quark and target state, one can define 16 Wigner distributions at the leading 
twist. All these Wigner distributions are calculated in light cone dressed quark model using Hamiltonian perturbation 
theory. In our model, we obtained 8 independent distributions and we illustrate here only 4 of them, a complete discussion
of all the leading twist Wigner distributions can be found in \cite{Jai16}. In Sec. \ref{plots} we give numerical results
and three-dimensional plots of the independent distributions in momentum space, impact 
parameter space and mixed space. Finally, we summarize our results in Sec. \ref{conclusion}. 
\section{Quark Wigner Distributions in Dressed Quark Model}\label{WD}
The Wigner distribution of quarks is defined as \cite{Meissner09, Lorce11}
\ba
\rho^{[\Gamma]} (b_{\perp},k_{\perp},x,s,s') = \int \frac{d^2 \Delta_{\perp}}{(2\pi)^2} e^{-i \Delta_{\perp}.b_{\perp}}
W^{[\Gamma]} (\Delta_{\perp},k_{\perp},x,s,s') 
\ea
where, ${\bs b}_\perp$ is  the impact parameter space conjugate to momentum transfer ($\Delta_\perp$) of a dressed 
quark in the transverse direction. 
GTMDs are defined through quark-quark correlator $W^{[\Gamma]}$ at a fixed light-front time as 
\ba
W^{[\Gamma]} (\Delta_{\perp},k_{\perp},x,s,s')
&=&\int \!\!\frac{dz^{-}d^{2} z_{\perp}}{2(2\pi)^3}e^{i k.z}
 \Big{\langle}p^{+},\frac{\Delta_{\perp}}{2},s \Big{|}
\overline{\psi}(-\frac{z}{2})\Omega \Gamma \psi(\frac{z}{2}) \Big{|}
p^{+},-\frac{\Delta_{\perp}}{2}, s'  \Big{\rangle }  \Big{|}_{z^{+}=0}
\ea
The initial and final dressed quark states are defined in the symmetric frame, with the 
longitudinal momentum as $p^+$, the transverse momentum transfer is  $\Delta_\perp$ and 
$s$, $s'$ are the helicities of initial and final target state. The average four momentum of the quark is $k$, 
with $k^+=x p^+$, here $x$ is the longitudinal momentum fraction of the parton. $\Omega$ 
is the gauge link which is required for color gauge invariance and 
is chosen to be unity.

The state of a dressed quark with momentum $p$ and fixed helicity $s$ can be written in terms of multiparton 
wavefunctions (LFWFs) as the expansion of the Fock state \cite{Hari99}
\ba
  \Big{| }p^{+},p_{\perp},s \Big{\rangle} &=& \Phi^{s}(p) b^{\dagger}_{s}(p) | 0 \rangle +
 \sum_{s_1 s_2} \int \frac{dp_1^{+}d^{2}p_1^{\perp}}{ \sqrt{16 \pi^3 p_1^{+}}}
 \int \frac{dp_2^{+}d^{2}p_2^{\perp}}{ \sqrt{16 \pi^3 p_2^{+}}} \sqrt{16 \pi^3 p^{+}}
 \delta^3(p-p_1-p_2) \nn \\[1.5ex] 
 &&\times\Phi^{s}_{s_1 s_2}(p;p_1,p_2) b^{\dagger}_{s_1}(p_1) 
 a^{\dagger}_{s_2}(p_2)  | 0 \rangle 
 \ea
$\Phi^{s}(p)$ is the single quark state and $\Phi^{s}_{s_1 s_2}(p;p_1,p_2)$ is the quark-gluon LFWF. $s_1$ and $s_2$ are helicities of quark and 
gluon respectively. $\Phi^{s}(p)$ gives the wavefunction normalization of the quark. $\Phi^{s}_{s_1 s_2}(p;p_1,p_2)$ 
gives the probability amplitude to find a bare quark (gluon) with momentum $p_1 (p_2)$ and helicity $s_1 (s_2)$ inside the dressed quark.
Using the Jacobi momenta
\ba k_{i}^{+} = x_{i}P^+ ~\text{and}~~~ k_{i}^{\perp} =q_{i}^{\perp} +  x_{i}P^{\perp} \hspace{1cm}
\text{so that}\hspace{1cm}\sum_i x_i=1,~~~~~\sum_iq_{i\perp}=0\ea 
the two particle LFWF can be written in terms of boost invariant LFWF as
\ba\sqrt{P^+}\Phi(p; p_1, p_2) = \Psi(x_{i},q_{i}^{\perp})\ea
Using the two particle LFWF \cite{Hari99} and 
two component formalism \cite{Zhang93}
at leading twist, one obtains only four Dirac operators $\Gamma=\{\gamma^+, \gamma^+\gamma^5,  i \sigma^{+1}\gamma^{5}, 
i \sigma^{+2}\gamma^{5}\}$ which corresponds to  Wigner distributions for unpolarized, 
longitudinally polarized and transversely polarized dressed quark.  
So the quark-quark correlator using two particle LFWFs for different polarization at twist-2 are given by 
\ba\label{U}
W^{[\gamma^+]} (\Delta_{\perp},k_{\perp},x,s,s') & =&\sum_{\lambda_1',\lambda_{1},\lambda_2} 
\Psi^{*s'}_{\lambda_{1}' \lambda_2}(x,q'^{\perp}) \chi_{\lambda_1'}^{\dagger} \chi_{\lambda_1}
\Psi^{s}_{\lambda_1 \lambda_2}(x,q^{\perp})\\
\label{L}
W^{[\gamma^+\gamma^{5}]} (\Delta_{\perp},k_{\perp},x,s,s') & =&\sum_{\lambda_1',\lambda_{1},\lambda_2} 
\Psi^{*s'}_{\lambda_{1}' \lambda_2}(x,q'^{\perp}) \chi_{\lambda_1'}^{\dagger}\sigma_3  \chi_{\lambda_1}
\Psi^{s}_{\lambda_1 \lambda_2}(x,q^{\perp})\\
\label{T}
W^{[i\sigma^{+j}\gamma^{5}]} (\Delta_{\perp},k_{\perp},x,s,s') & =&\sum_{\lambda_1',\lambda_{1},\lambda_2}  
\Psi^{*s'}_{\lambda_{1}' \lambda_2}(x,q'^{\perp}) \chi_{\lambda_1'}^{\dagger}\sigma_j  \chi_{\lambda_1}
\Psi^{s}_{\lambda_1 \lambda_2}(x,q^{\perp})
\ea
with $\sigma_i$ are the three Pauli matrices.
Eqs. (\ref{U}), (\ref{L}) and (\ref{T}) gives unpolarized,
longitudinally polarized and transversely polarized GTMDs in terms of LFWFs.
For various combinations of unpolarized (U), longitudinally polarized (L) 
and transversely polarized (T) target and quark states, the quark-quark correlators can be 
parametrized into 16 Wigner distributions \cite{Liu15} at leading twist. 
We denote Wigner distributions by $\rho_{\lambda,\lambda'}$, where $\lambda$ and $\lambda'$ 
represents polarization of the target state and quark respectively.
In this model, we have 10 independent Wigner distributions out of which $\rho^{j}_{TT}({\bs b}_\perp, {\bs k}_\perp,x)$ 
is zero with $j=1, 2$ and  $\rho_{LU}=\rho_{UL}$.  However, we discuss only four independent Wigner distributions 
but the details of all the 10 Wigner distributions can be looked in Ref \cite{Jai16} and in that we have classified \
Wigner distribution in terms of polarization of the target states.\\

\noindent The unpolarized Wigner distribution 
\ba
\rho_{UU}({\bs b}_\perp, {\bs k}_\perp,x)=\frac1{2}\Big[\rho^{[\gamma^+]}({\bs b}_\perp, {\bs k}_\perp,x,\hat{\bs e}_z)+\rho^{[\gamma^+]}({\bs b}_\perp, {\bs k}_\perp,x,-\hat{\bs e}_z)\Big]
\label{rhouu} 
\ea
The unpolarized-longitudinally polarized Wigner distribution
\ba
\rho_{UL}({\bs b}_\perp, {\bs k}_\perp,x)=\frac1{2}\Big[\rho^{[\gamma^+\gamma^5]}({\bs b}_\perp, {\bs k}_\perp,x,\hat{\bs e}_z)+\rho^{[\gamma^+\gamma^5]}({\bs b}_\perp, {\bs k}_\perp,x,-\hat{\bs e}_z)\Big]
\ea
The longitudinal Wigner distribution
\ba
\rho_{LL}({\bs b}_\perp, {\bs k}_\perp,x)=\frac1{2}\Big[\rho^{[\gamma^+\gamma^5]}({\bs b}_\perp, {\bs k}_\perp,x,\hat{\bs e}_z)-\rho^{[\gamma^+\gamma^5]}({\bs b}_\perp, {\bs k}_\perp,x,-\hat{\bs e}_z)\Big]
\ea
The transverse  Wigner distribution
\ba\rho_{TT}({\bs b}_\perp, {\bs k}_\perp,x)=\frac1{2}\delta_{ij}\Big[\rho^{[i \sigma^{+j}\gamma^5]}({\bs b}_\perp, {\bs k}_\perp,x,\hat{\bs e}_i)-\rho^{[i \sigma^{+j}\gamma^5]}({\bs b}_\perp, {\bs k}_\perp,x,-\hat{\bs e}_i)\Big]
\ea
The pretzelous  Wigner distribution 
\ba
\rho^{\perp}_{TT}({\bs b}_\perp, {\bs k}_\perp,x)= \frac1{2}\epsilon_{ij}\Big[\rho^{[i \sigma^{+j}\gamma^5]}({\bs b}_\perp, {\bs k}_\perp,x,\hat{\bs e}_i)-\rho^{[i \sigma^{+j}\gamma^5]}({\bs b}_\perp, {\bs k}_\perp,x,-\hat{\bs e}_i)\Big]
\label{rhotti}
\ea
Thus, the analytical expressions for the 
four Wigner distributions that we study are given by
\ba
\rho_{UU}({\bs b}_\perp, {\bs k}_\perp,x)&=& N \!\!\int\!\! d^2 \!\!\Delta_{\perp} \frac{\cos(\Delta_{\perp}b_{\perp})}{D(q_\perp)D(q'_\perp)}
 \Big[\frac{\Big(4k_\perp^2-\Delta_\perp^2 (1-x)^2\Big)(1+x^2)}{x^2(1-x)^3}+\frac{4 m^2(1-x)}{x^2}\Big]
\label{rhouu} \\
\rho_{UL}({\bs b}_\perp, {\bs k}_\perp,x)&=&N \int d^2 \Delta_{\perp} \frac{\sin(\Delta_{\perp}b_{\perp})}{D(q_\perp)D(q'_\perp)}
 \Big[\frac{4\Big(k_x\Delta_y-k_y\Delta_x\Big)(1+x)}{x^2(1-x)}\Big]\label{rhoul}\\
\rho_{LL}({\bs b}_\perp, {\bs k}_\perp,x)&=& N \!\!\int\!\! d^2 \Delta_{\perp} \frac{\cos(\Delta_{\perp}b_{\perp})}{D(q_\perp)D(q'_\perp)}
  \Big[\frac{\Big(4k_\perp^2-\Delta_\perp^2 (1-x)^2\Big)(1+x^2)}{x^2(1-x)^3}-\frac{4 m^2(1-x)}{x^2}\Big]\label{rholl}\\
\rho_{TT}({\bs b}_\perp, {\bs k}_\perp,x)&=&N \int d^2 \Delta_{\perp} \frac{\cos(\Delta_{\perp}b_{\perp})}{D(q_\perp)D(q'_\perp)}
  \Big[\frac{2\Big(4k_\perp^2-\Delta_\perp^2 (1-x)^2\Big)}{x(1-x)^3}\Big]\label{rhott}
\ea
where,\\
$$N=\frac{g^2 C_F}{2(2 \pi)^2},~~C_F \text{ is the color factor}$$
\ba 
D(q_\perp)=\Big{[ }   m^2 - \frac{m^2 + (k_\perp+\frac{\Delta_\perp(1-x)}{2})^2 }{x} - \frac{(k_\perp+\frac{\Delta_\perp(1-x)}{2})^2}{1-x} \Big]\nn\\
D(q'_\perp)=\Big{[ }   m^2 - \frac{m^2 + (k_\perp-\frac{\Delta_\perp(1-x)}{2})^2 }{x} - \frac{(k_\perp-\frac{\Delta_\perp(1-x)}{2})^2}{1-x} \Big]
\ea
We observed that, $\rho^{\perp}_{TT}({\bs b}_\perp, {\bs k}_\perp,x)=0$ in this model.
\section{Numericals and plots of probability densities in three dimension}\label{plots}
As mentioned in the  previous section, we study 4 independent Wigner distributions given in 
Eqs.(\ref{rhouu})\textendash(\ref{rhott}) which are five dimensional function of $b_x$, 
$b_y$, $k_x$, $k_y$ and $x$. We integrate over the momentum fraction $x$ 
which can take values only between $(0,1)$, in the dressed quark model. We observe the behavior
in the transverse momentum and impact parameter space of these 4 distributions.
One may also look at the mixed distributions by integrating out 
transverse momentum and transverse position along the perpendicular transverse directions.
In the mixed distribution one expects distribution to be real as the remaining variables are not 
restricted by Heisenberg's uncertainty principle. 
In this section, we show the numerical results and the plots of the quark Wigner distributions 
in transverse momentum space with $b_\perp =0.4~GeV^{-1}$, transverse position space with
${k}_\perp =0.4~GeV$ and mixed space. The integration limit of $\Delta_\perp$ should ideally 
be from zero to infinity.
\begin{figure}[h]
 \centering 
(a)\includegraphics[width=4.7cm,height=4cm]{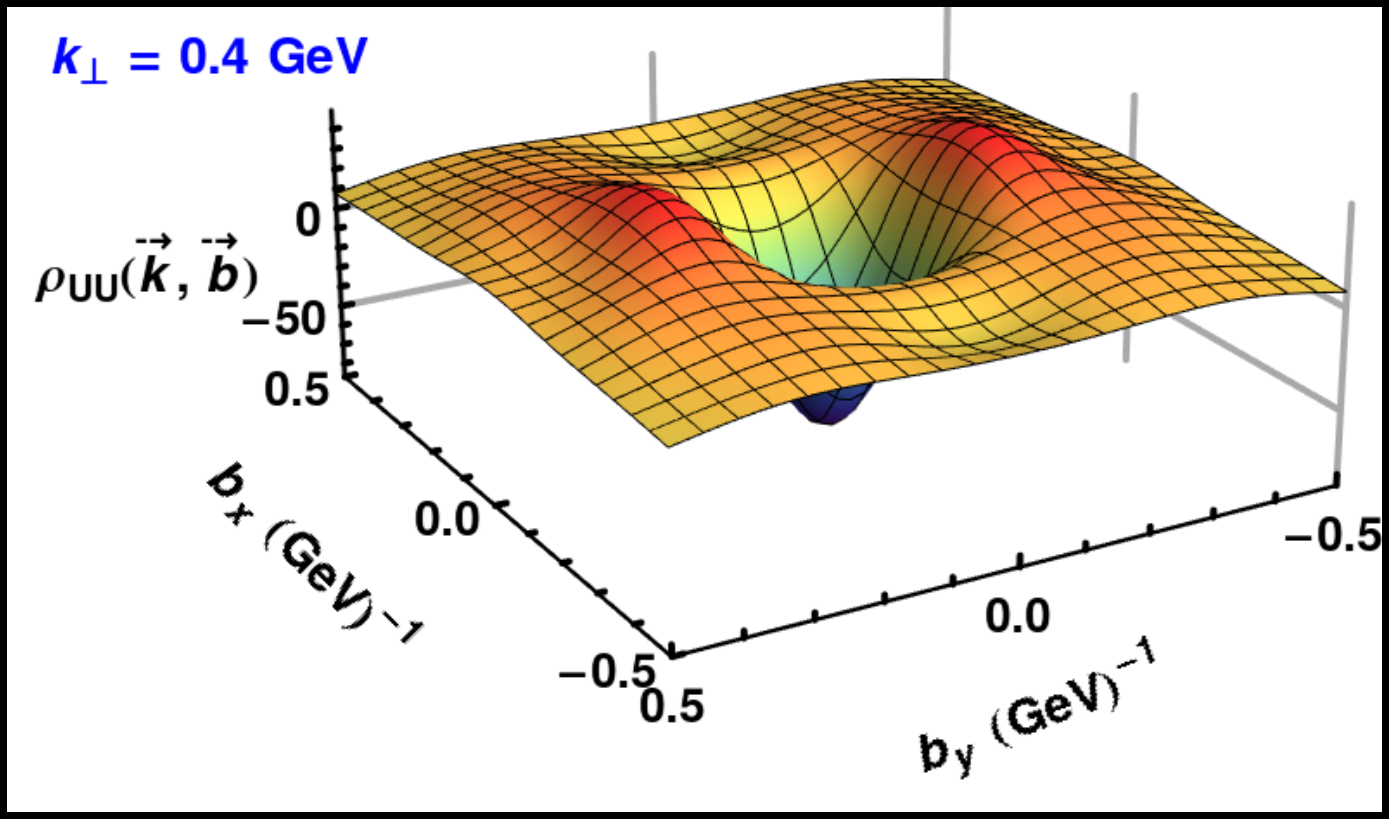} 
(b)\includegraphics[width=4.7cm,height=4cm]{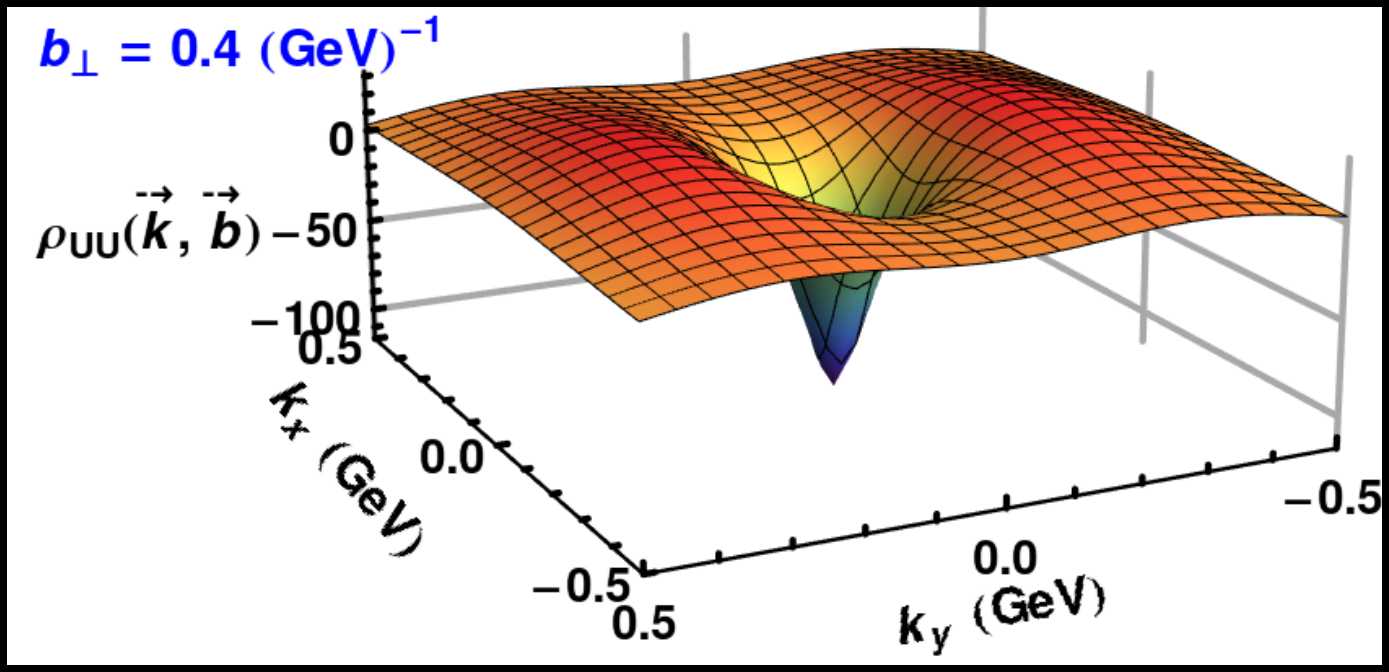}
(c)\includegraphics[width=4.7cm,height=4cm]{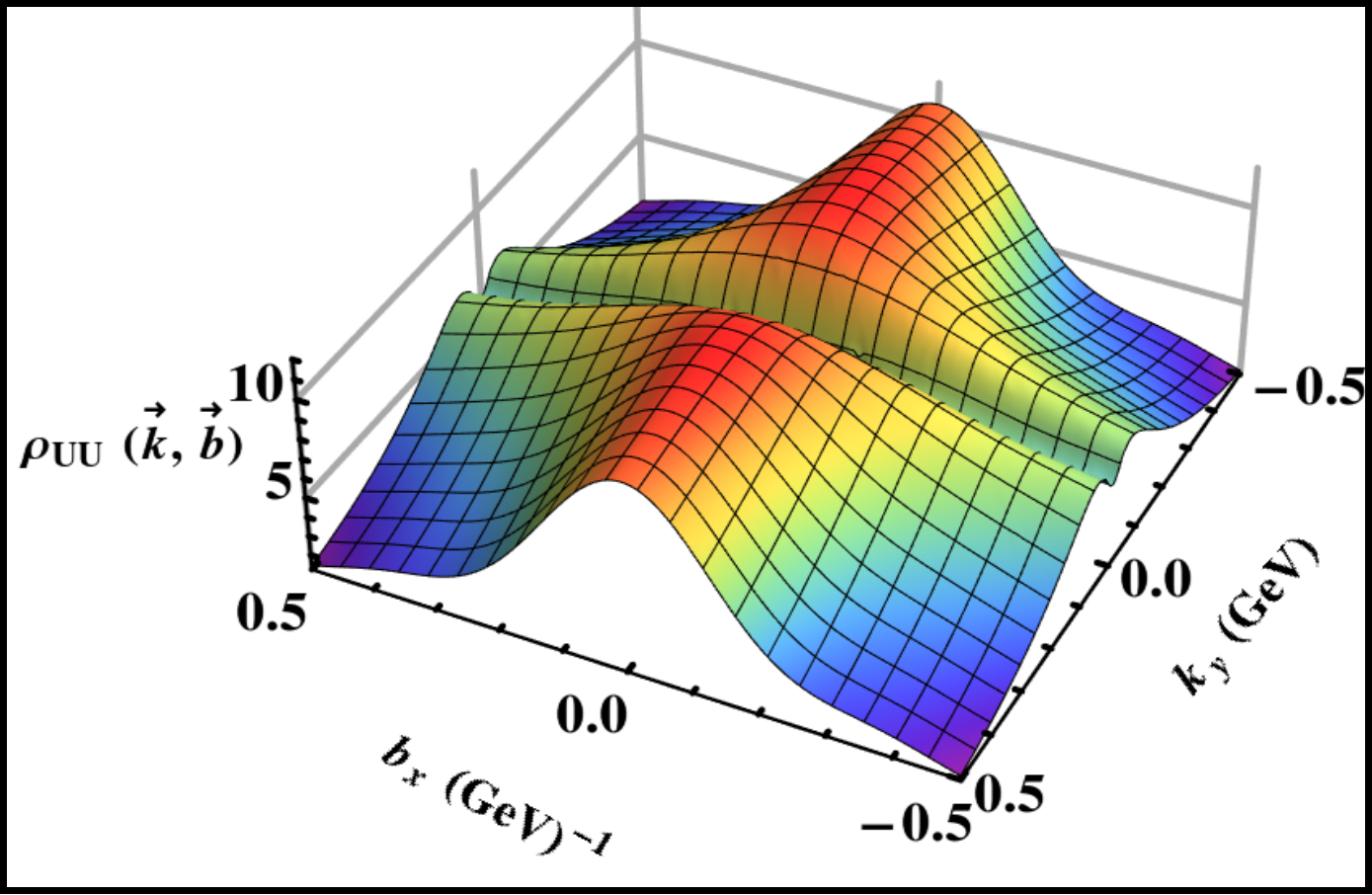}\\[2ex]
(d)\includegraphics[width=4.7cm,height=4cm]{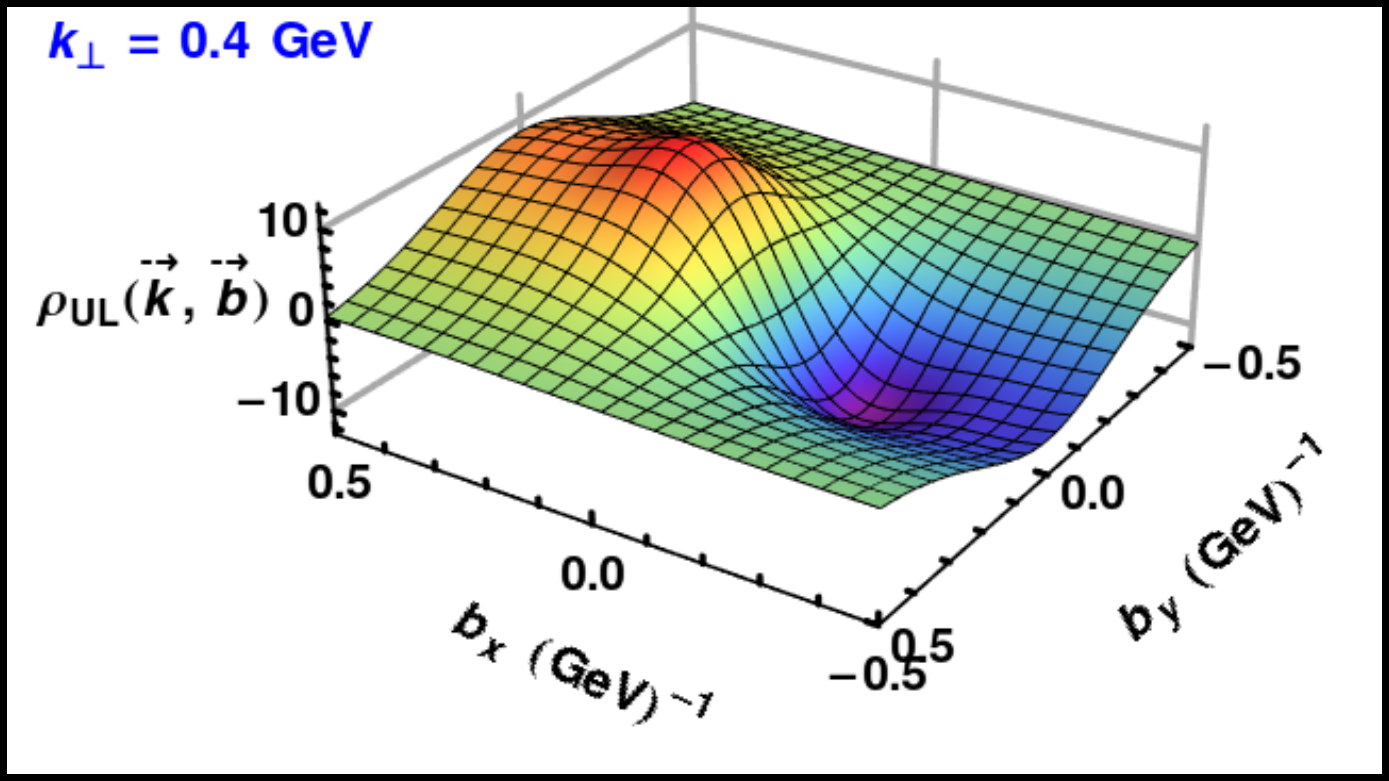}
(e)\includegraphics[width=4.7cm,height=4cm]{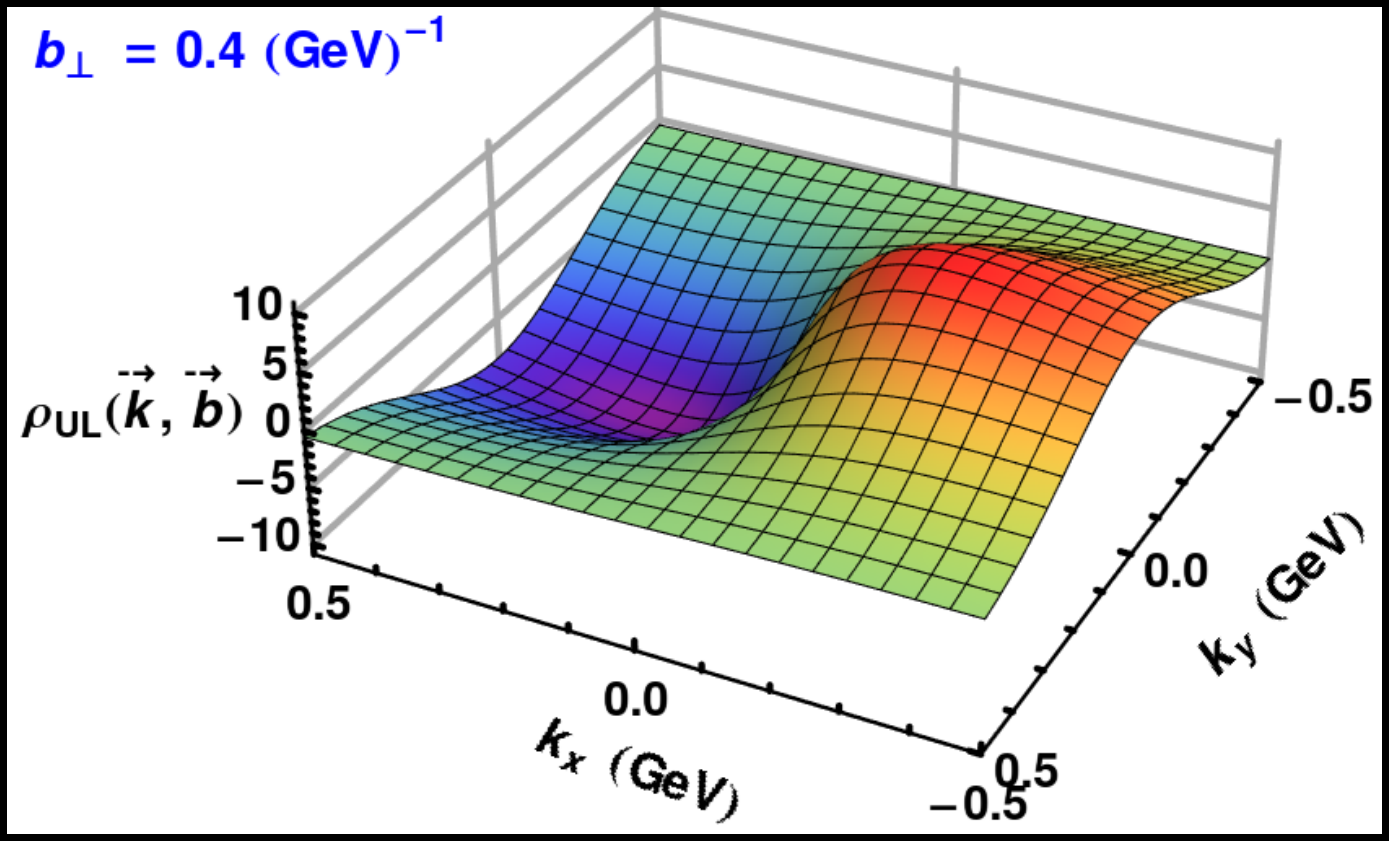}
(f)\includegraphics[width=4.7cm,height=4cm]{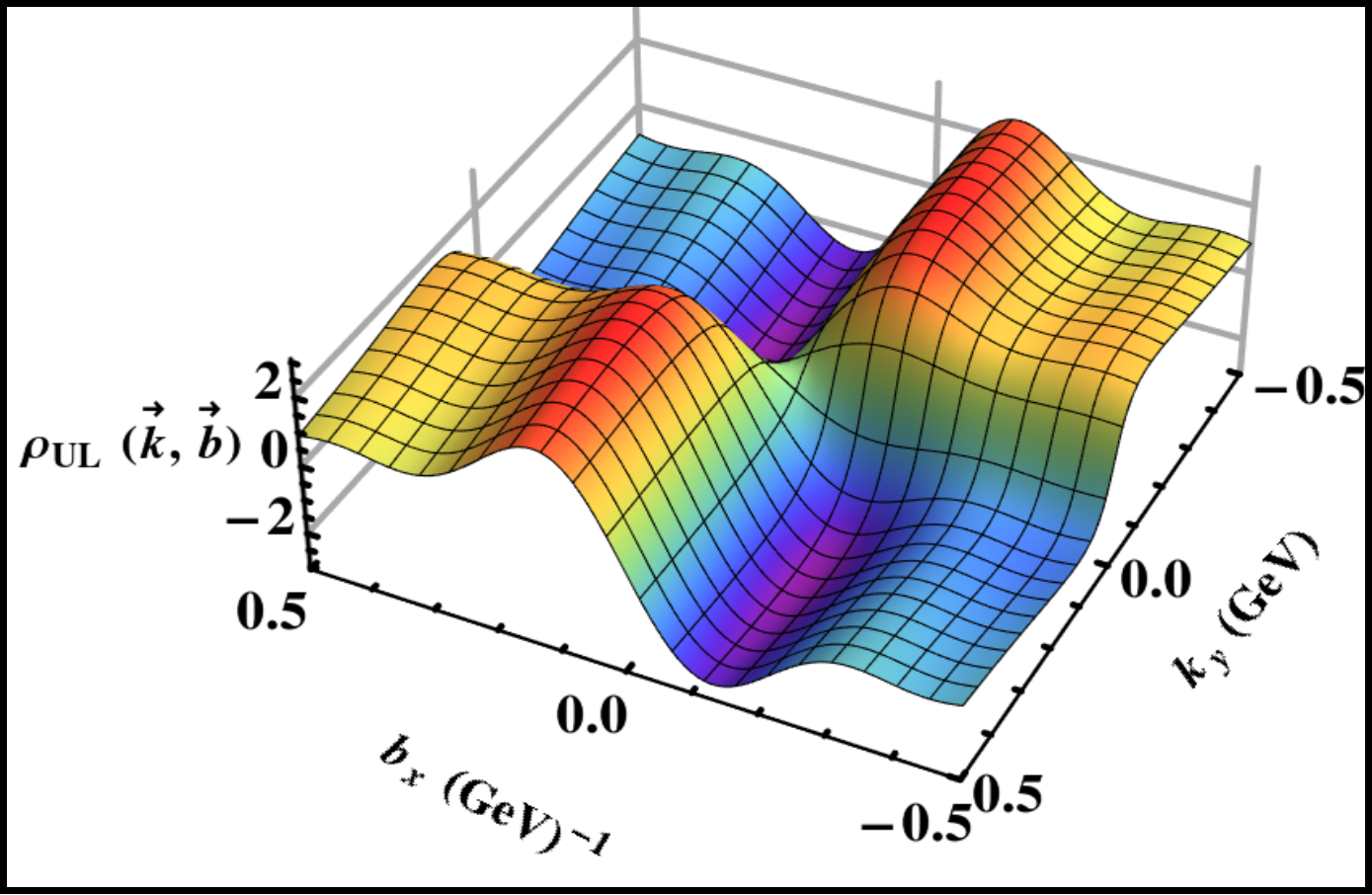}  
  \caption{3D plot of Wigner distributions $\rho_{UU}$ and $\rho_{UL}$ in $\boldmath {\bs k}_\perp-$space, 
  ${\bs b}_\perp-$space and $b_x-k_y$ space with $\Delta_{\perp max} = 20~GeV$}
  \label{rhou}
\end{figure}
However, in the numerical integration, we have used a cutoff on $\Delta_\perp$ integration, $\Delta_{\perp max}=20~GeV$.
It is to be noted that in the previous work \cite{Asmita14, Asmita15}, a lower value of $\Delta_{\perp max}$ was used and 
the results depend on this cutoff. We use Levin's method to perform numerical integration over $x$ and 
$\Delta_\perp$. It is a very effective method as it gives good convergence and the results do not depend on 
$\Delta_{\perp max}$.  
\begin{figure}[h]
 \centering 
(a)\includegraphics[width=4.7cm,height=4cm]{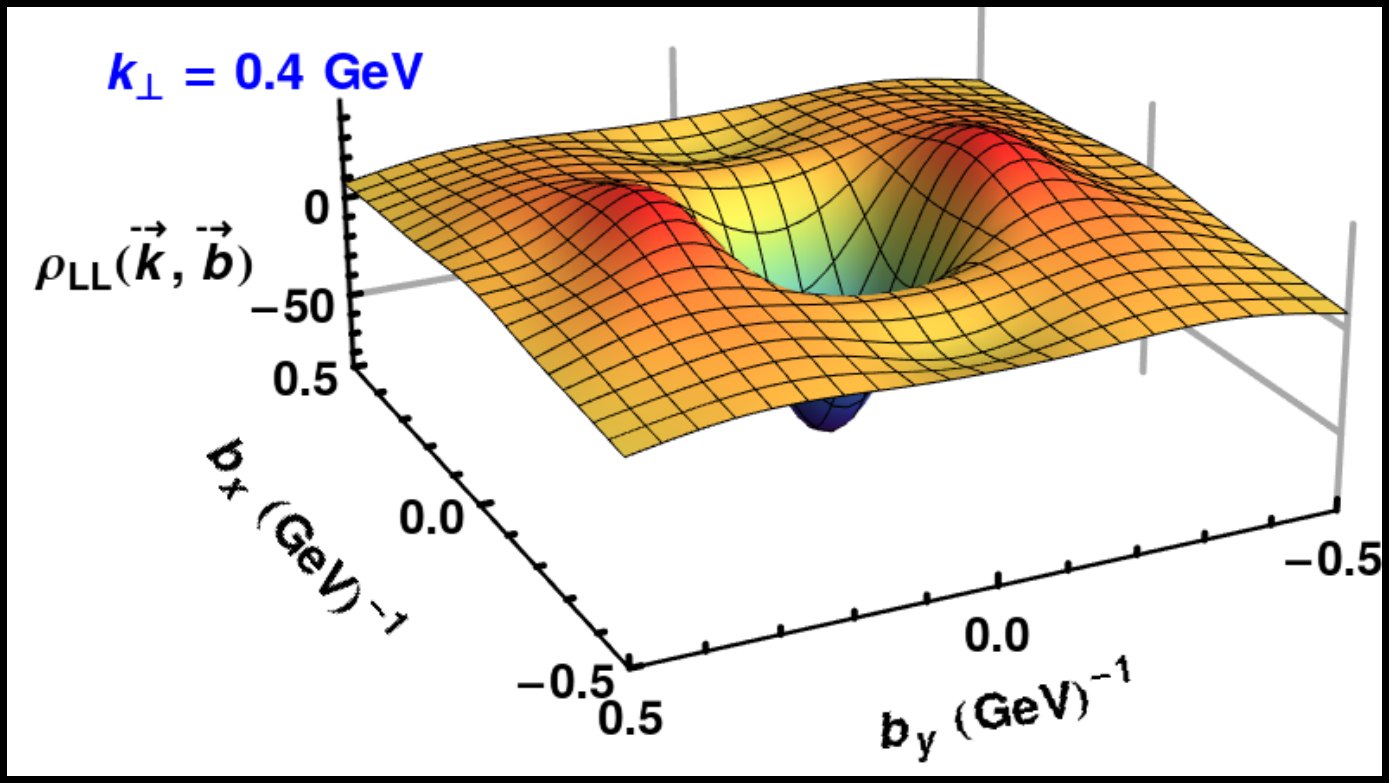}
(b)\includegraphics[width=4.7cm,height=4cm]{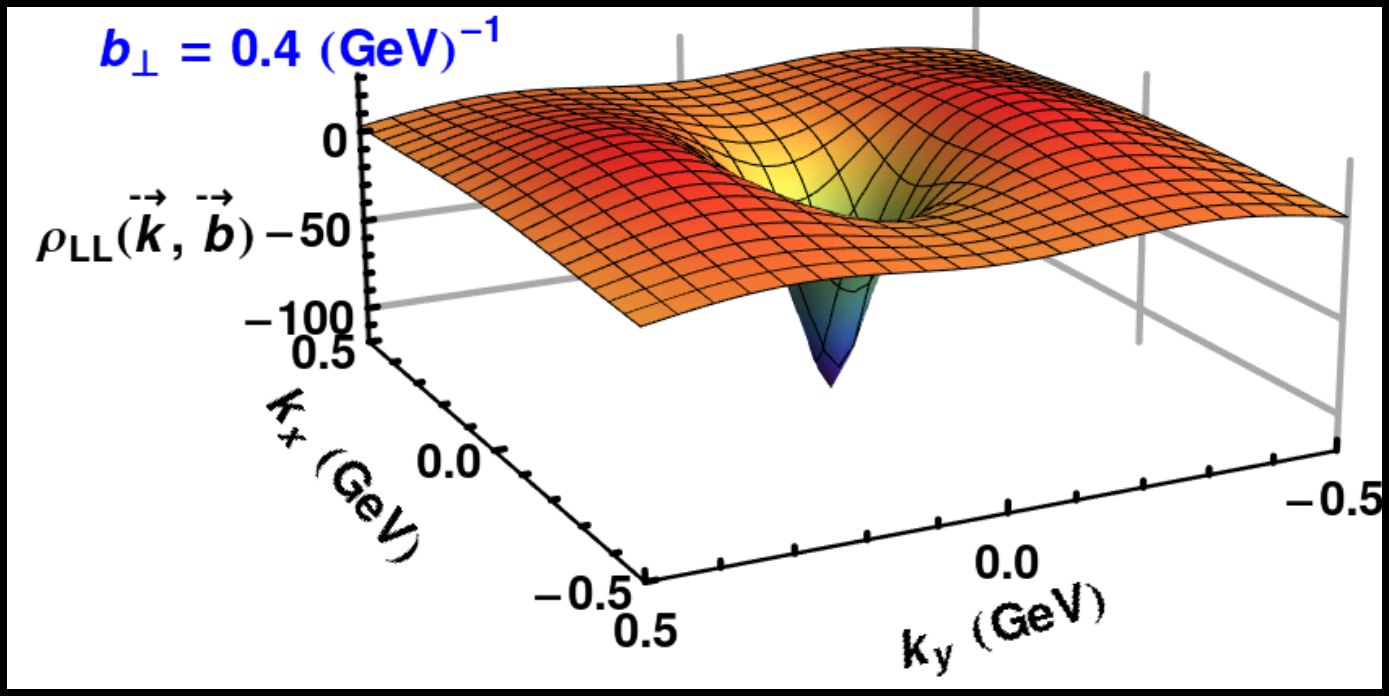}
(c)\includegraphics[width=4.7cm,height=4cm]{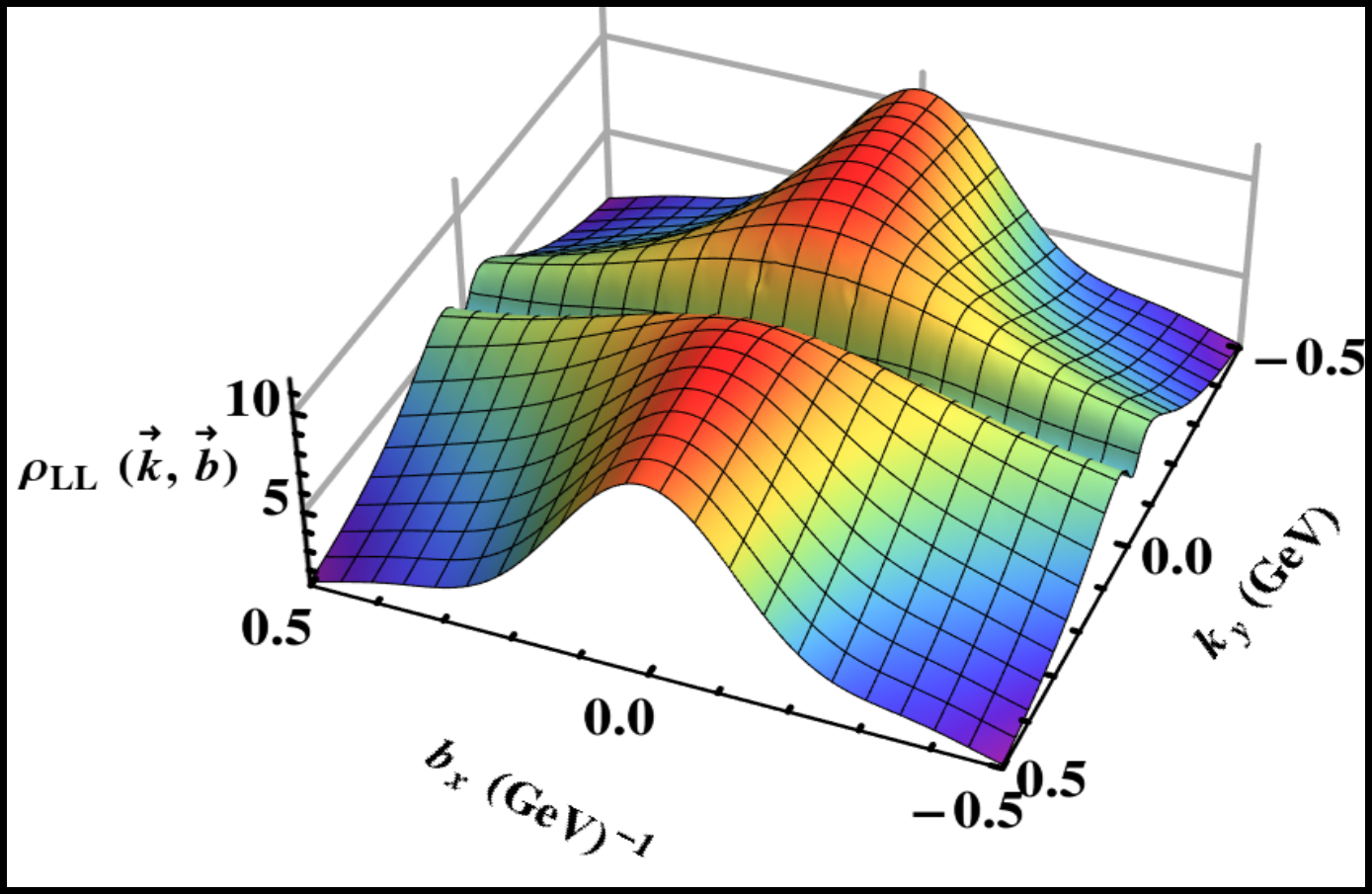}  \\[2ex]
(d)\includegraphics[width=4.7cm,height=4cm]{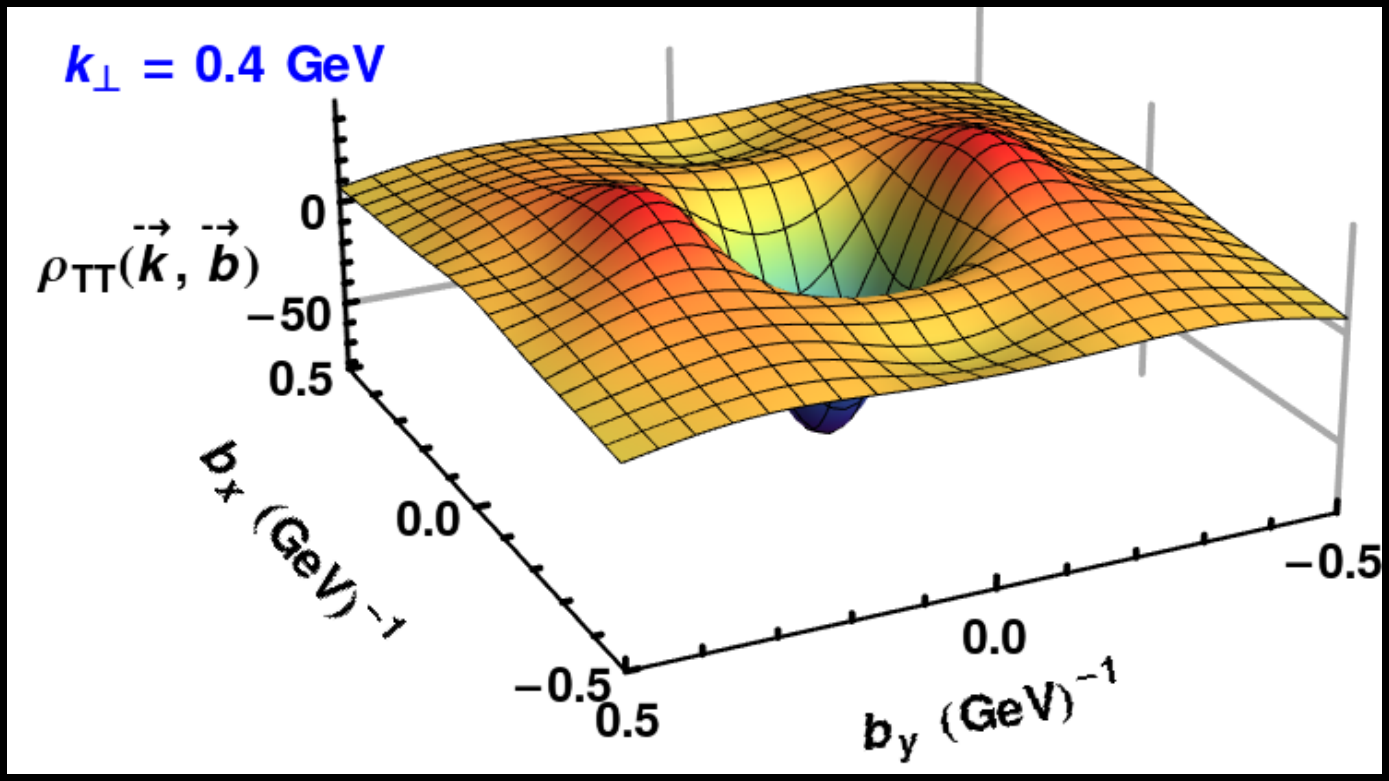}
(e)\includegraphics[width=4.7cm,height=4cm]{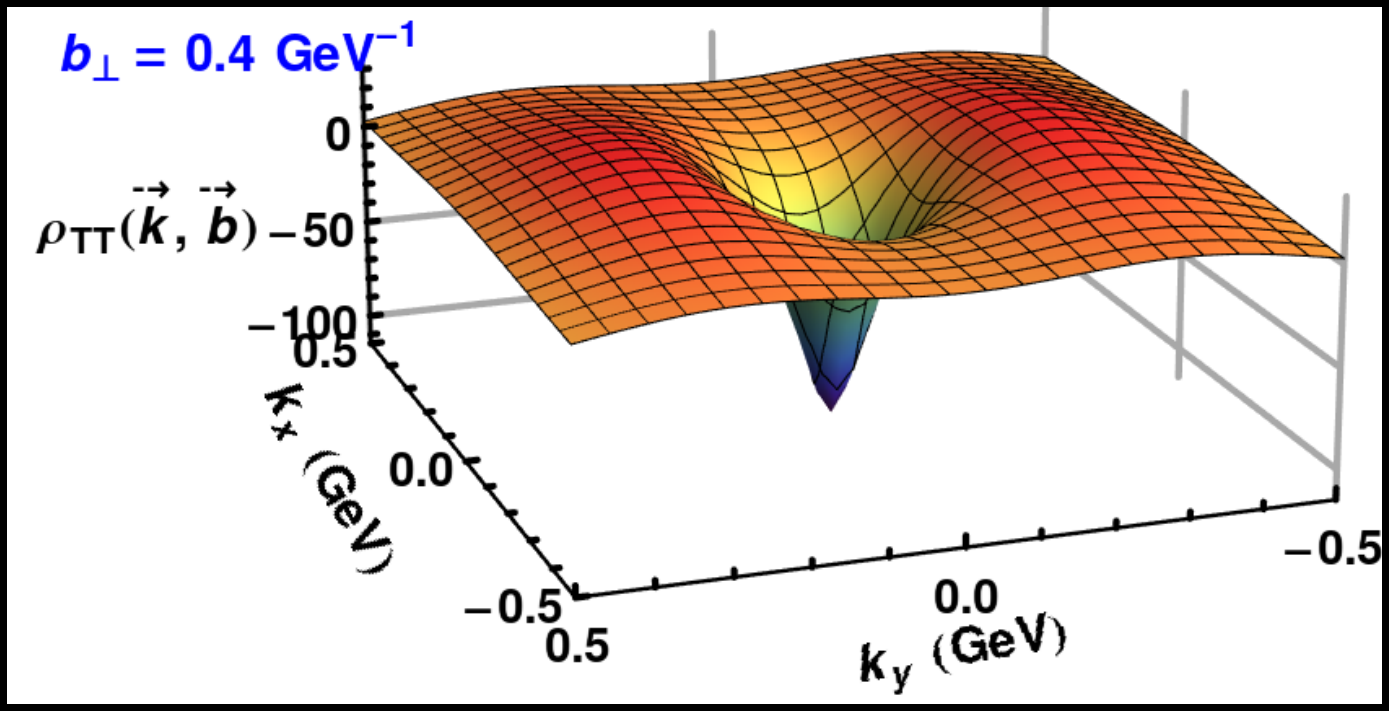}
 (f)\includegraphics[width=4.7cm,height=4cm]{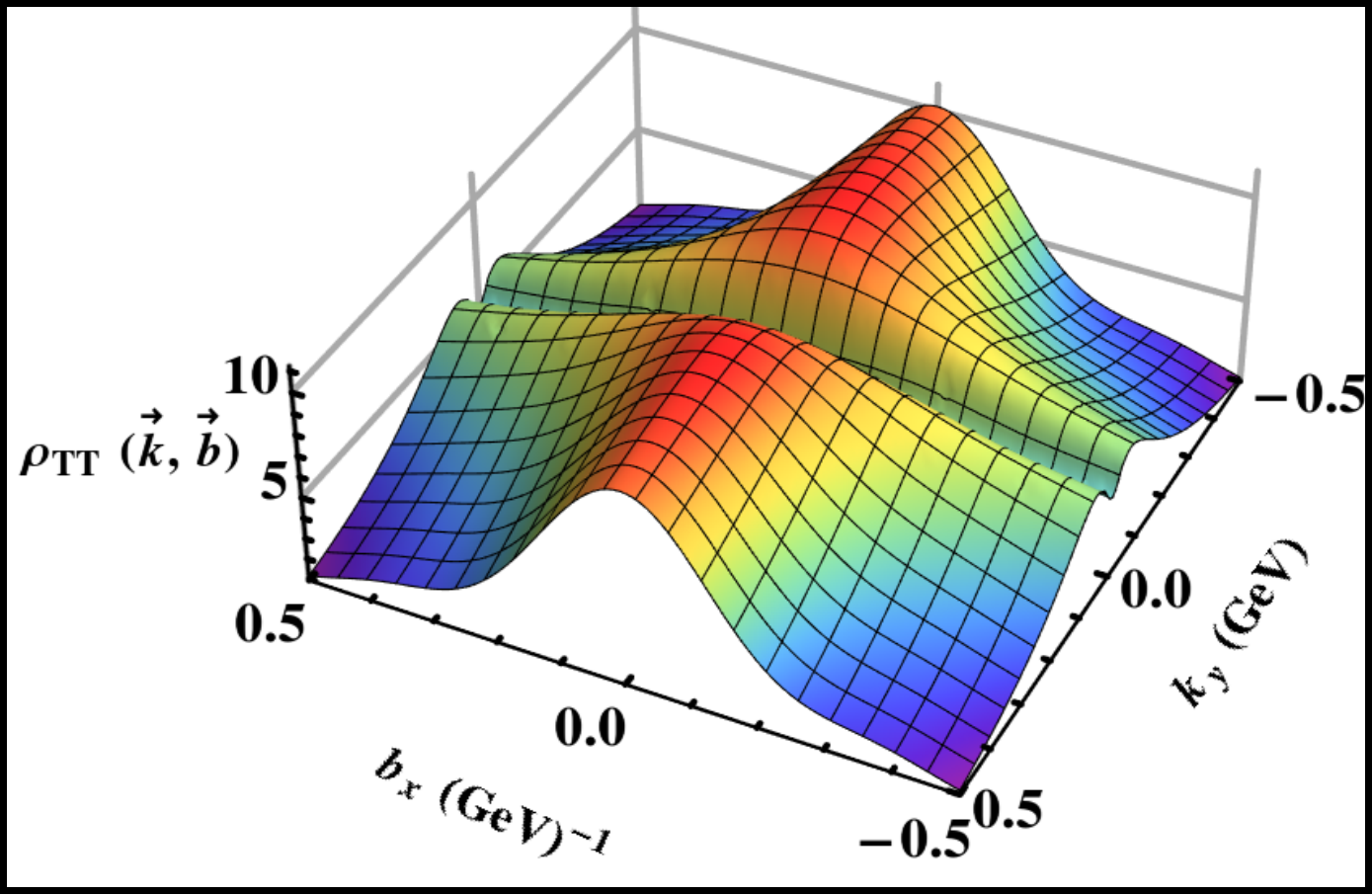}  
  \caption{3D plot of Wigner distribution $\rho_{LL}$ and $\rho_{TT}$ in ${\bs k}_\perp-$space, 
  ${\bs b}_\perp-$space and $b_x-k_y$ space with $\Delta_{\perp max} = 20~GeV$}
 \label{rhol}
\end{figure}

For numerical evaluation, in all the plots we have taken $m=0.33~$GeV and $\Delta_{\perp max} = 20~$ GeV.
Fig~\ref{rhou}(a) show a three-dimensional plot of Wigner distribution $\rho_{UU}$ in ${\bs k}_\perp-$space with 
$\bs{b}_\perp =0.4~GeV^{-1}~\hat{\bs e}_y$. $\rho_{UU}$ has a minimum and pick up negative value at $k_x=0$ and $k_y=0$. Fig~\ref{rhou}(b)
show the Wigner distribution $\rho_{UU}$ in ${\bs b}_\perp-$space with $\bs{k}_\perp =0.4~GeV~\hat{\bs e}_y$.
The behavior of ${\bs k}_\perp-$space is similar to ${\bs b}_\perp-$space with a different minima at $b_x=0$ and $b_y=0$ but it is still negative. Fig~\ref{rhou}(c) show the Wigner distribution $\rho_{UU}$ in mixed space where we have integrated out 
$k_x$ and $b_y$ dependence, thus we get the probability densities in the $b_x-k_y$ plane.

Fig.~\ref{rhou}(d) shows a three-dimensional plot of Wigner distributions $\rho_{UL}$ in momentum space with $\bs{b}_\perp =0.4~GeV^{-1}~\hat{\bs e}_y$. 
Fig.~\ref{rhou}(e) shows a three-dimensional plot of Wigner distributions $\rho_{UL}$ in impact parameter space with $\bs{k}_\perp =0.4~GeV~\hat{\bs e}_y$.
Figs.~\ref{rhou}(d) and (e) have similar nature. 
These two graphs show dipole structure as observed in Ref~\cite{Lorce11,Asmita14}. 
Fig.~\ref{rhou}(f) shows a three-dimensional plot of Wigner distributions
in mixed space which shows the quadruple structure.

Fig~\ref{rhol}(a)-(c) shows the three-dimensional plot of Wigner distributions $\rho_{LL}$ in ${\bs k}_\perp-$space, ${\bs b}_\perp-$space and the mixed space
respectively. Fig. \ref{rhol} for $\rho_{LL}$ shows similar nature as Fig~\ref{rhou} for $\rho_{UU}$ as expected.

Fig.~\ref{rhol}(d)-(f) shows the three-dimensional plot of Wigner distribution $\rho_{TT}$ in ${\bs k}_\perp-$space, 
${\bs b}_\perp-$space and mixed space respectively. In this case, both the quark and the target state are transversely polarized
in the direction say $x$-direction. It is important to note that nature of $\rho_{TT}$ is similar to 
$\rho_{UU}$ and $\rho_{LL}$.

\section{Conclusion}\label{conclusion}
 In this work, we include the transverse polarization of the quark and the target state  unlike in the previous work \cite{Asmita14}. 
 In this model, we study only 4 independent quark Wigner distributions as illustration and 
 the details will be given in Ref \cite{Jai16}. The unpolarized $\rho_{UU}$,  longitudinally polarized $\rho_{LL}$
and transversity distributions $\rho_{TT}$ show similar nature. $\rho_{LU}$ is equal to $\rho_{UL}$, which is related to orbital angular momentum
of the quark and the pretzelous Wigner distribution $\rho^{\perp}_{TT}$ is zero. We have used an improved method of numerical integration that gives better convergence and the results
are independent of $\Delta_{\perp max}$.
\begin{acknowledgements}
J.M. would like to thank Science and Engineering Research Board(SERB),  
for providing financial support and Light Cone organizers for their kind hospitality.
\end{acknowledgements}

%

\end{document}